\documentclass[12pt]{article}
\usepackage{latexsym,amssymb,amsmath}
\usepackage{stmaryrd}
\begin{document}
\baselineskip=16pt

\newcommand{\la}{\langle}
\newcommand{\ra}{\rangle}
\newcommand{\psp}{\vspace{0.4cm}}
\newcommand{\pse}{\vspace{0.2cm}}
\newcommand{\ptl}{\partial}
\newcommand{\dlt}{\delta}
\newcommand{\sgm}{\sigma}
\newcommand{\al}{\alpha}
\newcommand{\be}{\beta}
\newcommand{\G}{\Gamma}
\newcommand{\gm}{\gamma}
\newcommand{\vs}{\varsigma}
\newcommand{\Lmd}{\Lambda}
\newcommand{\lmd}{\lambda}
\newcommand{\td}{\tilde}
\newcommand{\vf}{\varphi}
\newcommand{\yt}{Y^{\nu}}
\newcommand{\wt}{\mbox{wt}\:}
\newcommand{\rd}{\mbox{Res}}
\newcommand{\ad}{\mbox{ad}}
\newcommand{\stl}{\stackrel}
\newcommand{\ol}{\overline}
\newcommand{\ul}{\underline}
\newcommand{\es}{\epsilon}
\newcommand{\sta}{\theta}
\newcommand{\dmd}{\diamond}
\newcommand{\clt}{\clubsuit}
\newcommand{\vt}{\vartheta}
\newcommand{\ves}{\varepsilon}
\newcommand{\dg}{\dagger}
\newcommand{\tr}{\mbox{Tr}}
\newcommand{\ga}{{\cal G}({\cal A})}
\newcommand{\hga}{\hat{\cal G}({\cal A})}
\newcommand{\Edo}{\mbox{End}\:}
\newcommand{\for}{\mbox{for}}
\newcommand{\kn}{\mbox{ker}}
\newcommand{\Dlt}{\Delta}
\newcommand{\rad}{\mbox{Rad}}
\newcommand{\rta}{\rightarrow}
\newcommand{\mbb}{\mathbb}
\newcommand{\lra}{\Longrightarrow}
\newcommand{\X}{{\cal X}}
\newcommand{\Y}{{\cal Y}}
\newcommand{\Z}{{\cal Z}}
\newcommand{\U}{{\cal U}}
\newcommand{\V}{{\cal V}}
\newcommand{\W}{{\cal W}}
\setlength{\unitlength}{3pt}

\begin{center}{\Large \bf  Moving-Frame Approach to Nonlinear
\\ Internal Waves in Oceans} 
\end{center}
\vspace{0.2cm}

\begin{center}{\large Xiaoping Xu}\end{center}
\begin{center}{Hua Loo-Keng Key Mathematical Laboratory\\Institute of Mathematics, Academy of Mathematics \& System Sciences\\
Chinese Academy of Sciences, Beijing 100190, P.R. China}
\footnote{Research supported
 by China NSF 10871193}\end{center}

\begin{abstract}{In this article, we introduce a moving-frame
approach to the geophysical equation of two-dimensional uniformly
stratified rotational fluid in oceans and find a family of exact
solutions containing ten arbitrary parameter
functions.}\end{abstract}

\section{Introduction}
The system of nonlinear equations
$$\Dlt\psi_t-g\rho_x-fv_z=\psi_x\Dlt\psi_z-\psi_z\Dlt\psi_x,\eqno(1.1)$$
$$v_t+f\psi_z=\psi_xv_z-\psi_zv_x,\eqno(1.2)$$
$$\rho_t+\frac{N^2}{g}\psi_x=\psi_x\rho_z-\psi_z\rho_x\eqno(1.3)$$
are used in geophysical fluid dynamics for investigating internal
waves in uniformly stratified incompressible fluids (oceans), where
$\psi, v, \rho$ are functions in $t,x,z$, $\Dlt=\ptl_x^2+\ptl_z^2$
is the two-dimensional Laplacian, $g$ is the gravitational
acceleration, $f$ is the Coriolis parameter and the quantity $N$
appears due to the density stratification of a fluid and is constant
 under the linear stratification hypothesis.

 Kistovich and Chashechkin (1991) [KC] used the system (1.1)-(1.3)
with $f=0$ to investigate two non-unidirectional wave beams
propagating and interacting in stratified fluid. Moreover, Lombard
and  Riley (1996) found an exact solution of the system (1.1)-(1.3)
with $f=0$ that describes stability of a single internal plane wave.
Using the system  with $f=0$,  Tabaei and Akylas (2003) [TA] study
certain nonlinear internal gravity wave beams. Furthermore, Tabaei,
Akylas and Lamb (2005) [TAL] studied the nonlinear effects in
reflecting and colliding internal wave beams. The above system with
$f\neq 0$ was used by Ibragimov [I] to model weakly nonlinear
interactions governing the time behavior of the oceanic energy
spectrum.

In [II1], N. Ibragimov and R. Ibragimov proved that the system
(1.1)-(1.3) is self-adjoint and obtained various conservation laws.
Using separation of variables, they also found some generalized
invariant solutions of the system. Invoking the software DYMSM 2.3,
 N. Ibragimov, R. Ibragimov and Kovalev [IIK] obtained the Lie point
 symmetries of the system (1.1)-(1.3) in terms of vector fields.
 Moreover, they found certain rotational invariant solutions.  Based
 on the maximal Lie subalgebra of the vector fields,  N. Ibragimov and R. Ibragimov
 [II2] got two additional conservation law.

In this article, we introduce a moving-frame approach to the system
(1.1)-(1.3), motivated from our earlier works on fluid equations
(cf. [X]) and find a family of exact solutions containing ten
arbitrary parameter functions. Our approach captures mote rotational
features of the fluid. The parameter functions also make our
solutions more applicable to practical models. In order to let more
people understand the symmetries of the system, we present intuitive
derivations of certain easily-using symmetry transformations, some
of which are used to obtain more general form of exact solutions.

\section{Intuitive  Symmetry Analysis}
Let $\al$ be a function in $t$. The transformation
$$\psi(t,x,z)\mapsto\psi(t,x+\al,z),\;v(t,x,z)\mapsto
v(t,x+\al,z),\;\rho(t,x,z)\mapsto\rho(t,x+\al,z) \eqno(2.1)$$
 changes the equations (1.1)-(1.3)
to:
$$\al'\Dlt\psi_x+\Dlt\psi_t-g\rho_x-fv_z=\psi_x\Dlt\psi_z-\psi_z\Dlt\psi_x,\eqno(2.2)$$
$$\al'v_x+v_t+f\psi_z=\psi_xv_z-\psi_zv_x,\eqno(2.3)$$
$$\al'\rho_x+\rho_t+\frac{N^2}{g}\psi_x=\psi_x\rho_z-\psi_z\rho_x,\eqno(2.4)$$
where the independent variable $x$ is replaced by $x+\al$ and the
subindices denote the partial derivatives with respect to the
original independent variables. Moreover, the transformation
$$\psi(t,x,z)\mapsto\psi(t,x,z)-\al'z,\;v(t,x,z)\mapsto
v(t,x,z),\;\rho(t,x,z)\mapsto\rho(t,x,z) \eqno(2.5)$$ changes the
equations (1.1)-(1.3) to:
$$\Dlt\psi_t-g\rho_x-fv_z=\psi_x\Dlt\psi_z-\psi_z\Dlt\psi_x+\al'\Dlt\psi_x,\eqno(2.6)$$
$$v_t+f\psi_z-f\al'=\psi_xv_z-\psi_zv_x+\al'v_x,\eqno(2.7)$$
$$\rho_t+\frac{N^2}{g}\psi_x=\psi_x\rho_z-\psi_z\rho_x+\al'\rho_x.\eqno(2.8)$$
Furthermore, the transformation
$$\psi(t,x,z)\mapsto\psi(t,x,z),\;v(t,x,z)\mapsto
v(t,x,z)+f\al,\;\rho(t,x,z)\mapsto\rho(t,x,z) \eqno(2.9)$$ leaves
the equations (1.1) and (1.3) invariant, and changes the equation
(1.2) to
$$f\al'+v_t+f\psi_z=\psi_xv_z-\psi_zv_x.\eqno(2.10)$$
Thus the transformation
$$T_{1,\al}(\psi(t,x,z))=\psi(t,x+\al,z)-\al'z,\;\;T_{1,\al}(v(t,x,z))=v(t,x+\al,z)+f\al,\eqno(2.11)$$
$$T_{1,\al}(\rho(t,x,z))=\rho(t,x+\al,z)\eqno(2.12)$$
is a symmetry transformation of the system (1.1)-(1.3). Similarly,
we have the following symmetry transformation
$$T_{2,\al}(\psi(t,x,z))=\psi(t,x,z+\al)+\al'x,\;\;T_{2,\al}(v(t,x,z))=v(t,x,z+\al),\eqno(2.13)$$
$$T_{2,\al}(\rho(t,x,z))=\rho(t,x,z+\al)-\frac{N^2}{g}\al.\eqno(2.14)$$
Since $\psi$ appears in the equations (1.1)-(1.3) with spacial
partial derivatives, we have the following symmetry of the equations
(1.1)-(1.3):
$$S_{\al,\be,\gm}(\psi(t,x,z))=\psi(t,x+\al,z+\be)+\be'x
-\al'z+\gm,\eqno(2.15)$$
$$S_{\al,\be,\gm}(v(t,x,z))=v(t,x+\al,z+\be)+f\al,\eqno(2.16)$$
$$S_{\al,\be,\gm}(\rho(t,x,z))=\rho(t,x+\al,z+\be)-\frac{N^2}{g}\be,\eqno(2.17)$$
 where $\al,\be,\gm$ are  any functions in $t$.

 Let us do degree
analysis. Take
$$\mbox{deg}\:\psi=\ell_1,\;\;\mbox{deg}\:x=\mbox{deg}\:z=\ell_2.\eqno(2.18)$$
To make all the terms in (1.1) having the same degree, we have to
take
$$\mbox{deg}\:\Dlt\psi_t=\mbox{deg}\:\psi_x\Dlt\psi_z\lra
\ell_1-2\ell_2-\mbox{deg}\:t=2\ell_1-4\ell_2\lra
\mbox{deg}\:t=2\ell_2-\ell_1,\eqno(2.19)$$
$$\mbox{deg}\:\rho_x=\mbox{deg}\:\psi_x\Dlt\psi_z\lra
-\ell_2+\mbox{deg}\:\rho=2\ell_1-4\ell_2\lra
\mbox{deg}\:\rho=2\ell_1-3\ell_2,\eqno(2.20)$$
$$\mbox{deg}\:v_z=\mbox{deg}\:\psi_x\Dlt\psi_z\lra
-\ell_2+\mbox{deg}\:v=2\ell_1-4\ell_2\lra
\mbox{deg}\:v=2\ell_1-3\ell_2.\eqno(2.21)$$ To make (1.2)
homogeneous, we take
$$\mbox{deg}\:v_t=\mbox{deg}\:\psi_z=\mbox{deg}\:\psi_xv_z\lra
\ell_1-\ell_2=3\ell_1-5\ell_2\lra\ell_1=2\ell_2.\eqno(2.22)$$ Under
the above assumption, (1.3) is homogeneous. In summary, we have
$$\mbox{deg}\:\psi=2\ell_2,\;\;\mbox{deg}\:\rho=\mbox{deg}\:v=\ell_2,\;\;
\mbox{deg}\:t=0.\eqno(2.23)$$ Since the equation (1.1)-(1.3) do not
contain variable coefficients, the are translation invariant.
Therefore we have the symmetry transformation:
$$T_{a,b}(\psi(t,x,z))=b^{-2}\psi(t+a,bx,bz),\;\;T_{a,b}(v(t,x,z))=b^{-1}v(t+a,bx,bz),\eqno(2.24)$$
$$T_{a,b}(\rho(t,x,z))=b^{-1}\rho(t+a,bx,bz),\eqno(2.25)$$ where
$a,b\in\mbb{R}$ such that $b\neq0$.

\section{Moving-Frame Approach}

Let $\gm$ be a function in $t$. Denote the  {\it moving frame}
$$\X=x\cos\gm+z\sin\gm,\qquad \Z=z\cos\gm-x\sin\gm.\eqno(3.1)$$
Then
$$\ptl_t(\X)=\gm'\Z,\qquad
\ptl_t(\Z)=-\gm'\X.\eqno(3.2)$$ By the chain rule of taking partial
derivatives,
$$\ptl_x=\cos\gm\;\ptl_\X-\sin\gm\;\ptl_\Z,\qquad
\ptl_z=\sin\gm\;\ptl_\X+\cos\gm\;\ptl_\Z.\eqno(3.3)$$
 Solving the above system, we get
$$\ptl_{\X}=\cos\gm\:\ptl_x+\sin\gm\:\ptl_z,\qquad
\ptl_{\Z}=-\sin\gm\:\ptl_x+\cos\gm\:\ptl_z.\eqno(3.4)$$ Moreover,
(3.1) and (3.4) imply
$$\ptl_\X(\Z)=0,\qquad\ptl_\Z(\X)=0.\eqno(3.5)$$
 In
particular,
$$\Dlt=\ptl_x^2+\ptl_z^2=\ptl_{\X}^2+\ptl_{\Z}^2,\;\;x^2+z^2=\X^2+\Z^2.\eqno(3.6)$$

Assume
$$\psi=\vt(t,x,z)+\xi(t,\X),\;\;v=\kappa(t,x,z)+\eta(t,\X),\;\;\rho=\varphi(t,x,z)+\zeta(t,\X),\eqno(3.7)$$
where $\xi,\eta,\zeta$ are function in $t,\X$, and
$\vt,\kappa,\varphi$ are functions in $t,x,z$ that $\vt$ is
quadratic and $\kappa,\varphi$ are linear in $x,z$.
 Now
$$\Dlt(\psi)=\Dlt\vt+\xi_{\X\X}\eqno(3.8)$$
by (3.6), and $\Dlt\vt$ is a function in $t$. Moreover,
$$\Dlt(\psi_t)=\Dlt\vt_t+\xi_{\X\X t}+\ptl_t(\X)\xi_{\X\X\X}=\Dlt\vt_t+\xi_{\X\X
t}+\gm'\Z\xi_{\X\X\X}\eqno(3.9)$$ by (3.2), and
$$\Dlt(\psi_x)=[\Dlt(\psi)]_x=\xi_{\X\X\X}\cos\gm,\;\Dlt(\psi_z)=[\Dlt(\psi)]_z=\xi_{\X\X\X}\sin\gm.\eqno(3.10)$$
Thus
\begin{eqnarray*}\psi_x\Dlt\psi_z-\psi_z\Dlt\psi_x&=&(\vt_x+\xi_\X\cos\gm)\xi_{\X\X\X}\sin\gm-(\vt_z+\xi_\X\sin\gm)\xi_{\X\X\X}\cos\gm
\\&=&(\vt_x\sin\gm-\vt_z\cos\gm)\xi_{\X\X\X}=-\ptl_\Z(\vt)\xi_{\X\X\X}\hspace{3.5cm}(3.11)\end{eqnarray*}
by (3.4).

Note that
\begin{eqnarray*}\psi_xv_z-\psi_zv_x&=&(\vt_x+\xi_\X\cos\gm)(\kappa_z+\eta_{_\X}\sin\gm)
-(\vt_z+\xi_\X\sin\gm)(\kappa_x+\eta_{_\X}\cos\gm)\\
&=&-\ptl_\Z(\vt)\eta_{_\X}+\ptl_\Z(\kappa)\xi_\X+\vt_x\kappa_z-\vt_z\kappa_x\hspace{5.2cm}(3.12)\end{eqnarray*}
by (3.4). Similarly,
$$\psi_x\rho_z-\psi_z\rho_x=-\ptl_\Z(\vt)\zeta_\X+\ptl_\Z(\vf)\xi_\X+\vt_x\vf_z-\vt_z\vf_x.\eqno(3.13)$$
Moreover,
$$v_t=\kappa_t+\eta_t+\gm'\Z\eta_\X,\;\;\rho_t=\vf_t+\zeta_t+\gm'\Z\zeta_\X.\eqno(3.14)$$

Now the equations  (1.1)-(1.3) become $$\Dlt\vt_t+\xi_{\X\X
t}+\gm'\Z\xi_{\X\X\X}-g(\vf_x+\eta_{_\X}\cos\gm)-f(\kappa_z+\zeta_\X\sin\gm)
=-\ptl_\Z(\vt)\xi_{\X\X\X},\eqno(3.15)$$
$$\kappa_t+\eta_t+\gm'\Z\eta_{_\X}+f(\vt_z+\xi_\X\sin\gm)=-\ptl_\Z(\vt)\eta_{_\X}+\ptl_\Z(\kappa)\xi_\X+\vt_x\kappa_z-\vt_z\kappa_x,\eqno(3.16)$$
$$\vf_t+\zeta_t+\gm'\Z\zeta_\X+\frac{N^2}{g}(\vt_x+\xi_\X\cos\gm)=-\ptl_\Z(\vt)\zeta_\X+\ptl_\Z(\vf)\xi_\X+\vt_x\vf_z-\vt_z\vf_x.\eqno(3.17)$$
We rewrite them as
$$\Dlt\vt_t-g\varphi_x-f\kappa_z+\xi_{\X\X
t}-g\eta_{_\X}\cos\gm-f\zeta_\X\sin\gm+(\gm'\Z+\ptl_\Z(\vt))\xi_{\X\X\X}
=0,\eqno(3.18)$$
$$\kappa_t+\eta_t+f\vt_z-\vt_x\kappa_z+\vt_z\kappa_x+(\gm'\Z+\ptl_\Z(\vt))\eta_{_\X}+(f\sin\gm-\ptl_\Z(\kappa))\xi_\X
=0,\eqno(3.19)$$
$$\vf_t+\zeta_t+\frac{N^2}{g}\vt_x-\vt_x\vf_z+\vt_z\vf_x+(\gm'\Z+\ptl_\Z(\vt))\zeta_\X+\left(\frac{N^2}{g}\cos\gm-\ptl_\Z(\vf)\right)\xi_\X
=0.\eqno(3.20)$$

In order to solve the above system, we assume
$$\vt=-\frac{\gm'}{2}\Z^2-\frac{(\al'\X+\be')\Z}{\al}+\frac{\al_1}{2}\X^2+\be_1\X\eqno(3.21)$$
for some functions $\al,\al_1,\be,\be_1$ in $t$, and
$$\Dlt\vt_t-g\varphi_x-f\kappa_z=0,\eqno(3.22)$$
$$\kappa_t+f\vt_z-\vt_x\kappa_z+\vt_z\kappa_x=0,\eqno(3.23)$$
$$\vf_t+\frac{N^2}{g}\vt_x-\vt_x\vf_z+\vt_z\vf_x=0.\eqno(3.24)$$

According to (3.1), (3.5) and (3.21),
$$\gm'\Z+\ptl_\Z(\vt)=-\frac{\al'\X+\be'}{\al},\eqno(3.25)$$
$$\vt_x=\left(\gm'\Z+\frac{\al'\X+\be'}{\al}\right)\sin\gm+\left(-\frac{\al'}{\al}\Z+\al_1\X+\be_1\right)\cos\gm,\eqno(3.25)$$
$$\vt_z=-\left(\gm'\Z+\frac{\al'\X+\be'}{\al}\right)\cos\gm+\left(-\frac{\al'}{\al}\Z+\al_1\X+\be_1\right)\sin\gm.\eqno(3.26)$$
In particular,
$$-\vt_x\ptl_z+\vt_z\ptl_x=-\left(\gm'\Z+\frac{\al'\X+\be'}{\al}\right)\ptl_\X+\left(\frac{\al'}{\al}\Z-\al_1\X-\be_1\right)\ptl_\Z\eqno(3.27)$$
This motivates us to assume
$$\kappa=\al_2\X+\be_2\Z+\nu_2,\qquad\vf=\al_3\X+\be_3\Z+\nu_3,\eqno(3.28)$$
where $\al_2,\al_3,\be_3,\be_3,\nu_2,\nu_3$ are functions in $t$ to
be determined. Then (3.22)-(3.24) become
$$f(\al_3\cos\gm-\be_3\sin\gm)+g(\al_2\sin\gm+\be_2\cos\gm)=\al_1'-{\gm'}',\eqno(3.29)$$
\begin{eqnarray*}&
&(\al_2'-\be_2\gm')\X+(\be_2'+\al_2\gm')\Z-\left(\gm'\Z+\frac{\al'\X+\be'}{\al}\right)
\al_2+\left(\frac{\al'}{\al}\Z-\al_1\X-\be_1\right)\be_2
\\ & &+f\left[-\left(\gm'\Z+\frac{\al'\X+\be'}{\al}\right)\cos\gm+\left(-\frac{\al'}{\al}\Z+\al_1\X+\be_1\right)\sin\gm\right]
+\nu_2\:'=0,\hspace{1.1cm}(3.30)\end{eqnarray*}
\begin{eqnarray*}&
&(\al_3'-\be_3\gm')\X+(\be_3'+\al_3\gm')\Z-\left(\gm'\Z+\frac{\al'\X+\be'}{\al}\right)\al_3+\left(\frac{\al'}{\al}\Z
-\al_1\X-\be_1\right)\be_3
\\ & &+\frac{N^2}{g}\left[\left(\gm'\Z+\frac{\al'\X+\be'}{\al}\right)\sin\gm+\left(-\frac{\al'}{\al}\Z+\al_1\X+\be_1\right)\cos\gm\right]
+\nu_3\:'=0.\hspace{1.1cm}(3.31)\end{eqnarray*}

Observe that (3.30) is equivalent to
$$\al_2'-\frac{\al'}{\al}\al_2-(\al_1+\gm')\be_2+f\left(-\frac{\al'}{\al}\cos\gm+\al_1\sin\gm\right)=0,\eqno(3.32)$$
$$\be_2'+\frac{\al'}{\al}\be_2-f\left(\frac{\al'}{\al}\sin\gm+\gm'\cos\gm\right)=0,\eqno(3.33)$$
$$\nu_2\:'-\frac{(\al_2+f\cos\gm)\be'}{\al}+(f\sin\gm-\be_2)\be_1=0.\eqno(3.34)$$
By (3.33), we take
$$\be_2=f\sin\gm.\eqno(3.35)$$
Substituting it to (3.32), we get
$$\al_2'-\frac{\al'}{\al}\al_2-f\left(\frac{\al'}{\al}\cos\gm+\gm'\sin\gm\right)=0.\eqno(3.36)$$
So we take
$$\al_2=-f\cos\gm.\eqno(3.37)$$
According to (3.31), (3.35) and (3.37), we take $\nu_2=0$.

Next (3.31) is equivalent to
$$\al_3'-\frac{\al'}{\al}\al_3-(\al_1+\gm')\be_3+\frac{N^2}{g}\left(\al_1\cos\gm+\frac{\al'}{\al}\sin\gm\right)=0,\eqno(3.38)$$
$$\be_3'+\frac{\al'}{\al}\be_3+\frac{N^2}{g}\left(\gm'\sin\gm-\frac{\al'}{\al}\cos\gm\right)=0,\eqno(3.39)$$
$$\nu_3'-\left(\al_3-\frac{N^2}{g}\sin\gm\right)\frac{\be'}{\al}+\left(\frac{N^2}{g}\cos\gm-\be_3\right)\be_1=0.\eqno(3.40)$$
Similarly, we have the solutions:
$$\be_3=\frac{N^2}{g}\cos\gm,\;\;\al_3=\frac{N^2}{g}\sin\gm,\;\;\nu_3=0.\eqno(3.41)$$
Now (3.29) becomes $\al_1'-{\gm'}'=0$. So we take
$$\al_1=\gm'.\eqno(3.42)$$

Observe that (3.18)-(3.20) become
$$\xi_{\X\X
t}-g\eta_{_\X}\cos\gm-f\zeta_\X\sin\gm-\frac{\al'\X+\be'}{\al}\xi_{\X\X\X}
=0,\eqno(3.43)$$
$$\eta_t-\frac{\al'\X+\be'}{\al}\eta_{_\X}
=0,\eqno(3.44)$$
$$\zeta_t-\frac{\al'\X+\be'}{\al}\zeta_\X
=0.\eqno(3.45)$$ By (3.44) and (3.45), we have
$$\eta=\phi'(\al\X+\be),\qquad\zeta=\mu'(\al\X+\be)\eqno(3.46)$$
for some one-variable functions $\phi$ and $\mu$. Moreover, (3.43)
yields
$$\xi=\frac{h(\al\X+\be)}{\al^2}+\int\frac{f\phi(\al\X+\be)\sin\gm+g\mu(\al\X+\be)\cos\gm}{\al}dt\eqno(3.47)$$
for some  one-variable function $h$, where $\X$ and $t$ should be
treated as independent variables in the integral.

By (3.1),
$$\Z\sin\gm-\X\cos\al=-x,\;\;\Z\cos\gm+\X\sin\al=z.\eqno(3.48)$$
 In summary, we have:\psp

{\bf Theorem 3.1}. {\it Let $\al,\be,\be_1,\gm,\sta_1,\sta_2,\sta_3$
be any differentiable functions in $t$ such that $\al\not\equiv 0$,
and let $\phi,\mu,h$ be any differentiable one-variable functions.
Denote $\X=x\cos\gm+z\sin\gm$ and $\Z=-x\sin\gm+z\cos\gm$. We have
the following solution of the equation (1.1)-(1.3):
\begin{eqnarray*}\hspace{2cm}\psi&=&\frac{\gm'}{2}(x^2+z^2)-\frac{(\al'\X+\be')\Z}{\al}+\be_1\X
+\frac{h(\al\X+\be)}{\al^2}\\
&
&+\int\frac{f\phi(\al\X+\be)\sin\gm+g\mu(\al\X+\be)\cos\gm}{\al}dt,\hspace{3.4cm}(3.49)\end{eqnarray*}
$$v=-fx+\phi'(\al\X+\be),\qquad\rho=\frac{N^2z}{g}+\mu'(\al\X+\be).\eqno(3.50)$$
Applying the symmetry transformation $S_{\sta_1,\sta_2,\sta_3}$ to
the above solution, we get a solution of the equation (1.1)-(1.3)
with ten arbitrary parameter functions. }

\vspace{0.7cm}

\noindent{\Large \bf References}

\hspace{0.5cm}

\begin{description}

\item[{[KC]}] A. V. Kistovich and Y. D. Chashechkin, Nonlinear
interactions of two-dimensional packets of monochromatic internal
waves, {\it Izv. Atmos. Ocean. Phys.} {\bf 27} (1991), 946-951.

\item[{[LR]}] P. N. Lombard and J. Riley, On breakdown into
turbulence propagating internal waves, {\it Dyn. Atmos. Oceans} {\bf
23} (1996), 345-355.

\item[{[TA]}] A. Tabaei and T. R. Akylas, Nonlinear internal gravity
wave beams, {\it J. Fluid Mech.} {\bf 482} (2003), 141-161.

\item[{[TAL]}] A. Tabaei, T. R. Akylas and K. G. Lamb, Nonlinear
effects in reflecting and colliding internal wave beams, {\it J.
Fluid Mech.} {\bf 526} (2005), 217-243.

\item[{[I]}] N. H. Ibragimov, Latitude-dependent classification of
singularities for resonant interaction between discrete-mode
internal gravity waves, {\it J. Phys. Oceanography}, to appear.

\item[{[II1]}] N. H. Ibragimov and R. N. Ibragimov,  Group analysis of nonlinear
internal waves in oceans I. self-adjointness, conservation laws and
invariant solutions, {\it Archives of ALGA} {\bf 6} (2009), 19-44.

\item[{[IIK]}] N. H. Ibragimov, R. N. Ibragimov and V. F. Kovalev,  Group analysis of nonlinear
internal waves in oceans II. the symmetries and rotationally
invariant solutions, {\it Archives of ALGA} {\bf 6} (2009), 45-54.

\item[{[II2]}] N. H. Ibragimov and R. N. Ibragimov,  Group analysis of nonlinear
internal waves in oceans III. additional conservation laws, {\it
Archives of ALGA} {\bf 6} (2009), 55-62.

\item[{[X]}] X. Xu, {\it Algebraic Approaches to Partial
Differential Equations}, Springer, Heidelberg/New York/Dordrecht/
London, 2013.

\end{description}

 \end{document}